\documentclass[aps,superscriptaddress,showpacs]{revtex4}
\usepackage{epsfig,graphics,amssymb,amsmath,subeqnarray}
\def\di{\displaystyle}\def\p{\partial}\def\u{{\bf u}}\def\A{{\bf A}}\def\e{{\bf e}}\def\btau{\boldsymbol{\tau}}\def\bgam{\boldsymbol{\gamma}}\def\gamd{\dot\bgam}\def\De{{\rm De}}
\def\f{{\bf f}}\def\n{{\bf n}}\def\t{{\bf t}}\def\dgamd{\stackrel{\triangledown}{\gamd}}
\def\bs{{\boldsymbol{\sigma}}}\def\dbtau{\stackrel{\triangledown}{\btau}}\def\B{{\bf B}}\def\h{{\bf h}}
\def\varepsilon{\epsilon}

\begin{document}

\title{Flapping motion and force generation in a viscoelastic fluid}
\author{Thibaud Normand}
\affiliation{D\'epartement de M\'ecanique, Ecole Polytechnique, 91128 Palaiseau Cedex, France.}
\author{Eric Lauga\footnote{Corresponding author. Email: elauga@ucsd.edu}}
\affiliation{Department of Mechanical and Aerospace Engineering, 
University of California San Diego,
9500 Gillman Drive, La Jolla CA 92093-0411, USA.}
\date{\today}
\begin{abstract}

In a variety of biological situations, swimming cells have to move through complex fluids. Similarly, mucociliary clearance involves the transport of polymeric fluids by beating cilia. 
Here, we consider the extent to which complex fluids could be exploited for force generation on small scales.  We consider a prototypical reciprocal motion (i.e. identical under time-reversal symmetry): The periodic flapping of a tethered   semi-infinite plane. In the Newtonian limit, such motion cannot be used for force generation according to Purcell's scallop theorem. In a polymeric fluid (Oldroyd-B, and its generalization), we show that this is not the case and  calculate explicitly the forces on the flapper for  small-amplitude sinusoidal motion. Three  setups are considered: a flapper near a wall, a flapper in a wedge, and a two-dimensional scallop-like flapper. In all cases, we show that at quadratic  order in the oscillation amplitude, the tethered flapping motion induces net  forces but no average flow. Our results demonstrate therefore that the  scallop theorem is not valid in polymeric fluids. The reciprocal component of the movement of biological appendages such as cilia can thus generate nontrivial forces in polymeric fluid such as mucus, and  normal-stress differences can be exploited as a pure viscoelastic force generation and propulsion method.

\end{abstract}

\maketitle

\section{Introduction}

Locomotion is a subject with a rich history, and with relevance  to all length scales in biology, not only in the animal kingdom - from large cetaceans to  spermatozoa - but also to bacteria, protozoa, and algae \cite{gray68,alexander03}. 

A particularly active area of research concerns  cell motility in viscous fluids. Classical work in the 1950-70's have elucidated the fundamental principles of fluid-based propulsion on small length scales 
\cite{lighthill75,lighthill76,brennen77,childress81,braybook}, but a number of physical issues are still unresolved. Most notably, the extent to which its surrounding environment plays an important role in the swimming of an individual cells is not well understood.

One relevant example is the locomotion of cells in complex fluids, such as the swimming of spermatozoa in cervical mucus \cite{fauci06,suarez06}. Cervical mucus is a high-viscosity cross-linked polymeric gel \cite{yudin89} located at the entrance of the uterus along the female reproductive tract. Its most notable feature is to possess relaxation time scales in the 1-100 seconds range \cite{wolf77_1,wolf77_2,wolf77_3,wolf78,wolf80,tam80}, which are much longer than the typical time scales involved in the beating of flagella (typical frequencies in the tens of hertz). As a consequence, the mechanics of locomotion in cervical mucus is significantly different  from locomotion in Newtonian flows. Experimentally,  spermatozoa swimming in complex fluid display a change in their beat pattern (smaller amplitude, wavelength), an increase in their beat frequency, and (as a result) a change of  their  swimming kinematics 
\cite{shukla78,katz78,katz80,katz81,rikmenspoel84,ishijima86,suarez92}. Theoretically, recent work predicted that cells 
actuating their flagella in a wave-like fashion should systematically swim  slower in a polymeric fluid than in a Newtonian fluid \cite{lauga07,Fu08,Fu_swimming}.

From a general standpoint, however, there is no reason to expect that going from a Newtonian to a viscoelastic fluid should systematically lead to a degradation of the swimming performance. Since a viscoelastic fluid displays nonlinear  rheological behavior \cite{birdvol1,birdvol2,doi88,larson99}, in general it should be possible for cells to exploit these nonlinearities to gain propulsive advantage. The goal of this paper is consider the simplest possible body deformation able to take advantage of these nonlinearities for force generation.

In a Newtonian flow, it is known that there exists a class of body deformations - those where the sequence of body shapes are identical under time-reversal symmetry, called ``reciprocal'' - that lead to exactly zero net propulsion in the Newtonian limit in the absence of inertia \cite{purcell77,childress04,vandenberghe04,alben05,lauga_purcell}. Such result is known as  Purcell's scallop theorem  \cite{purcell77}, and is due to the linearity and time-reversibility of Stokes equation of motion. The prototypical  reciprocal motion has only one degree of freedom, such as a flapping back-and-forth motion (e.g.  the  flapping of a wing, or the motion of  a scallop). As a result of the constraints of the scallop theorem, swimming microorganisms  are observed to always deform their body or their flagella in a wave-like fashion in order to move. Similarly, mucociliary clearance in our lungs involve fluid transport  by arrays of cilia beating periodically in a non-reciprocal manner  \cite{sleigh88}.

In this paper,  we consider tethered flapping motion in a variety of two-dimensional  settings in the absence of inertia as the simplest model of  biological reciprocal motion. We calculate the flow field perturbatively in the flapping amplitude for a polymeric fluid (Oldroyd-B and generalized models).  Although no net flow is obtained at second order for a periodic actuation, we show that net forces are generated. Reciprocal motions are therefore able to  harness normal stress differences  to generate net forces, and the scallop theorem breaks down in polymeric fluids. An implication of these results is the possibility for the reciprocal component of biological appendages, such as cilia, to generate nontrivial forces when moving in polymeric fluids such as mucus.

In many ways, our calculations are reminiscent of  classical work by GI Taylor and HK Moffatt on similarity solutions for  the Newtonian paint scraper problem \cite{taylor_scraper, batchelor_book,moffatt00}, and two-dimensional viscous vortices  \cite{moffatt64,moffatt80}. They are also related to the study of polymeric stress on steady (or ``acoustic'') streaming \cite{chang74,james77,rosenblat78,bohme92,bagchi66,frater67,frater68,goldstein75,chang77,chang79}, which is the physical phenomenon by which the periodic actuation of a solid object is able to exploit inertial forces to generate net flows. In particular, polymeric stresses lead to a reversal of the net flow direction \cite{chang74,james77,rosenblat78,bohme92}. Here, we are concerned however with purely viscoelastic forces, and ignore the influence of inertia. For work related to the current study in the limit of weakly viscoelastic flows (small Deborah number), see Ref.~\cite{Roper07}.

The paper is organized as follows. The general setup for our two-dimensional flapping calculations is presented in \S\ref{sectiontwo}.  We then calculate in detail the time-averaged forces on a semi-infinite flapper near a wall, and more generally in a wedge, for an Oldroyd-B fluid in \S\ref{sec_single}. The forces on a two-dimensional scallop are next calculated in \S\ref{sec_scallop}. We generalize our results, and discuss their implications  for biological propulsion in \S\ref{discussion}.


\section{Viscoelastic forces in small-amplitude  flapping: General results} \label{sectiontwo}

\subsection{Setup}
\label{setup}

\begin{figure}
\begin{center}
 \includegraphics[width=0.8\textwidth]{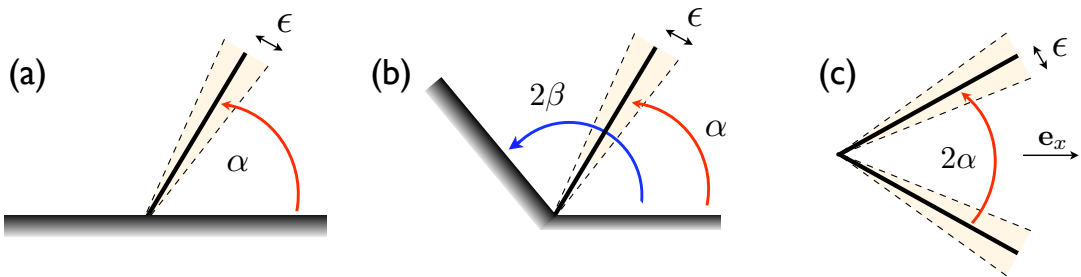}
\end{center}
 \caption{(color online) Cases considered in this paper: (a) two dimensional flapping motion near a no-slip surface (flapping of amplitude $\epsilon$ around a mean angle $\alpha$); (b) flapping motion in a no-slip wedge (wedge angle, $2\beta$); (c) scallop-like flapping motion (mean angle $2\alpha$ between the two arms of the scallop). In all cases, the hinge point of the flappers are fixed in the lab frame.}
\label{flapper}
\end{figure}

We consider a semi-infinite two-dimensional plane flapping with small amplitude in a viscoelastic fluid. The three setups we will consider are illustrated in Fig.~\ref{flapper}: (a) a flapper near a flat wall, (b) a flapper in a wedge, and (c) a scallop-like body made of two symmetric flappers. We will first calculate the flow on one half of the flow domain (e.g. $0\leq \theta \leq \alpha$ in Fig.~\ref{flapper}a) and deduce the flow in the second part of the domain (e.g.  $\alpha\leq \theta \leq \pi$ in Fig.~\ref{flapper}a) by analogy.

\subsection{Constitutive equation}
\label{mechanics}

We assume in this paper incompressible flow in the absence of inertia. 
The velocity field  is denoted $\u$ and the pressure field $p$.
In the fluid domain, Cauchy's equation of motion is therefore written as
\begin{subeqnarray}\label{cauchy}
\nabla\cdot \u & = & 0,\\
\nabla p &= & \nabla \cdot \btau,
\end{subeqnarray}
where $\btau$ is the deviatoric stress tensor. We first consider polymeric fluids described by the Oldroyd-B constitutive relationship \cite{birdvol1,birdvol2,larson99}
\begin{equation}\label{oldroyd}
\btau +\lambda_1 \dbtau= \eta [\gamd + \lambda_2 \dgamd ],
\end{equation}
with $\gamd=\nabla \u + \nabla \u ^T$.
In Eq.~\eqref{oldroyd} we have defined, for a tensor $\bf A$, the upper-convected derivative
$\stackrel{\triangledown}{{\bf A}} = \frac{\p {\bf A}}{\p t} + {\bf u}\cdot \nabla{\bf A} - (\nabla \u^T\cdot {\bf A} + {\bf A}\cdot \nabla \u)$;  $\lambda_1$ and $\lambda_2<\lambda_1$ are the  relaxation and retardation times of the fluid respectively. We consider periodic flapping motion with frequency $\omega$. We  non-dimensionalize rates and stresses in  Eqs.~\eqref{cauchy}-\eqref{oldroyd} by $\omega$ and $\eta \omega$ respectively, and lengths by some (arbitrary) length scale along the flapper. The dimensionless equations are therefore given by
\begin{subeqnarray}\label{oldroydB}
\nabla\cdot \u & = & 0,\\
\nabla p & =& \nabla \cdot \btau,\slabel{cauchy_2} \\
\btau +\De_1 \dbtau & = & \gamd + \De_2 \dgamd,
\end{subeqnarray}
where $\De_1=\lambda_1 \omega$ and $\De_2=\lambda_2 \omega$ are the two Deborah numbers for the flow, and where we use the same symbols for convenience. The Oldroyd-B constitutive relationship, which can be derived analytically from a dilute solution of perfectly elastic dumbbells \cite{larson99}, correctly models polymeric fluids up to order one Deborah numbers. For larger values of $\De$, more accurate models are necessary to capture the correct extensional rheology of polymeric flows. One of such models is considered in the discussion (FENE-P), and  we obtain the same results (see also \cite{lauga07}). Other Oldroyd-like models with more complex rheological characteristics are also discussed in \S\ref{discussion}.

\begin{figure}
\begin{center}
 \includegraphics[width=0.4\textwidth]{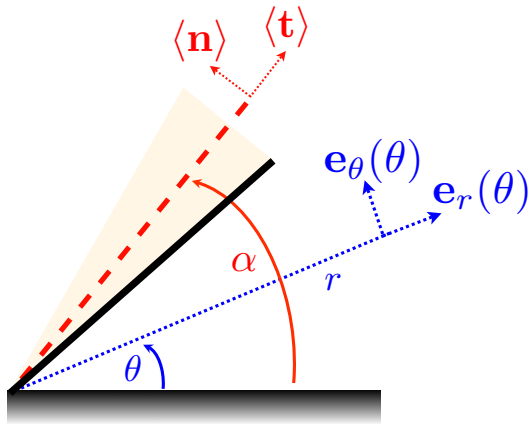}
\end{center}
 \caption{(color online) Setup and notation for the flapping calculation: the infinite flapper (black) has an opening angle  which oscillates in time around a mean value, $\alpha$ (red). For a point in the fluid domain at a distance $r$ from the hinge point,  the polar vectors are denoted $\e_r (\theta)$ and $\e_\theta(\theta)$  (blue). The polar vectors attached to the average flapper position are denoted  $\langle {\bf t} \rangle$ and $\langle {\bf n} \rangle$ (red). }
\label{notation}
\end{figure}

\subsection{Streamfunction}
\label{streamfunction}

Given the geometrical setup, it is most convenient to use polar coordinates, and  introduce the streamfunction, $\Psi$, defined by $u_{r}=({\partial\Psi}/{\partial\theta})/r$, $ u_{\theta}=-{\partial\Psi}/{\partial r}$.  The polar 
vectors  are denoted $\e_r  (\theta)$ and $\e_\theta (\theta)$ (see Fig.~\ref{notation}), and the velocity field is denoted $\u=u_r \e_r + u_\theta\e_\theta$.

\subsection{Perturbation expansion}
\label{expansion}

The calculations in the paper are done perturbatively in the amplitude of the flapping. We denote the instantaneous position of the flapper as $\theta = \alpha + \epsilon \Theta(t)$, with $\Theta(t)=\cos t$ an order-one oscillatory function, and perform the calculation in the asymptotic limit $\epsilon \ll 1$ using domain perturbation. We look therefore to solve the problem using an expansion of the form
\begin{eqnarray}\label{exp}
\{\mathbf{u},\Psi,\btau,p,\bs\}&=& 
\epsilon\{\mathbf{u}_{1},\Psi_1 ,{\btau}_{1},p_1, \bs_1 \} \\
&+& \epsilon^2\{\mathbf{u}_{2}, \Psi_2,{\btau}_{2},p_2, \bs_2 \}+...,\nonumber
\end{eqnarray}
where $\bs=-p{\bf 1} + \btau$ is the total stress tensor, and where all variables in in Eq.~\eqref{exp} are defined in the averaged domain $0\leq \theta \leq \alpha$.

\subsection{Instantaneous vs. average positions}
\label{average}

Since we perform a domain-perturbation expansion, we have to pay close attention to the distinction between instantaneous and average geometry. We  use planar polar coordinates. The vectors $\e_r$ and $\e_\theta$ on the flapper are functions of the azimuthal angle which oscillates in time: $\e_r(\alpha+\epsilon \Theta(t))$, $\e_\theta(\alpha+\epsilon \Theta(t))$. In the paper, we use the notations
$\langle {\bf t}\rangle = \e_r(\alpha)$ and $\langle {\bf n} \rangle = \e_\theta(\alpha)$ to denote the average directions along and perpendicular to the flapper axis.

\subsection{Fourier notations}
\label{Fourier}

It is easier to solve for the flow field using Fourier notations. The actuation is defined as $\Theta(t)=\Re\{e^{it}\}$, $\dot \Theta(t)=\Re\{i e^{it}\}$.  Since quadratic nonlinearities arise from boundary conditions and the constitutive model (see below), we anticipate a Fourier decomposition of the velocity field of the form 
\begin{subeqnarray}\label{fourier}
\u_1 & = &\Re \{ \tilde\u_1e^{it} \}, \\
\u_2 & = &\Re \{\tilde\u_2^{(0)} + \tilde\u_2^{(2)}e^{2it}   \},
\end{subeqnarray}
and we use similar notations for all other vector and scalar fields in the problem.

\subsection{General equations}
\label{general}

We now look in detail at the first two orders of Eq.~\eqref{oldroydB}. At leading order, the nonlinear terms disappear, and we obtain 
\begin{equation}
 \btau_{1}+\text{De}_{1}\frac{\partial\btau_{1}}{\partial t}=\gamd_{1}+\text{De}_{2}\frac{\partial\gamd_{1}}{\partial t},
\end{equation}
which becomes, in Fourier notations,
\begin{equation}\label{tau1_Fourier}
\tilde\btau_1 = \frac{1+i\De_2}{1+i\De_1}\tilde\gamd_1\cdot
\end{equation}
We then take the divergence of Eq.~\eqref{tau1_Fourier} and the curl of Eq.~\eqref{cauchy_2}, to obtain the equation for the  as streamfunction:
\begin{equation}\label{order1}
 \left(1+\frac{\partial}{\partial t}\text{De}_{2}\right)\nabla^{4}\Psi_{1}=0.
\end{equation}

At order $O(\epsilon^2)$, Eq.~\eqref{oldroydB} becomes
\begin{eqnarray}\label{order2}
&& \left(1+\frac{\partial}{\partial t}\text{De}_{1}\right)\btau_{2}
-\left(1+\frac{\partial}{\partial t}\text{De}_{2}\right)\gamd_{2}\quad \\
 &  
& \quad =\text{De}_{2}\left[\mathbf{u}_{1}\cdot\nabla\gamd_{1}-\left(^{t}\nabla\mathbf{u}_{1}\cdot\gamd_{1}+\gamd_{1}\cdot\nabla\mathbf{u}_{1}\right)\right] \nonumber\\
 &  &\quad\quad - \text{De}_{1}\left[\mathbf{u}_{1}\cdot\nabla\btau_{1}-\left(^{t}\nabla\mathbf{u}_{1}\cdot\btau_{1}+\btau_{1}\cdot\nabla\mathbf{u}_{1}\right)\right]. \nonumber \end{eqnarray}
Using Eq.~\eqref{tau1_Fourier}, the average of Eq.~\eqref{order2} becomes, using Fourier notations,
\begin{eqnarray}\label{order2_tocalculate}
  \tilde\btau_2^{(0)}-\tilde\gamd_2^{(0)}&=&\frac{1}{2}\frac{\text{De}_{2}-\text{De}_{1}}{1+i\text{De}_{1}}
  \left[\tilde\u_1^{*}\cdot\nabla\tilde\gamd_1\right. \\ && \left. -\left(^{t}\nabla\tilde\u_1^{*}\cdot\tilde\gamd_1+\tilde\gamd_1\cdot\nabla\tilde\u_1^{*}\right)\right], \nonumber
\end{eqnarray}
which can be evaluated using knowledge of the solution at order $O(\epsilon)$.

\subsection{Boundary conditions}
\label{bc}
Associated with the equations derived in \S\ref{general} and \S\ref{averagestresses}, we have to obtain boundary conditions at each order.  On steady surfaces, the boundary conditions can be simply stated. For the steady no-slip surfaces 
in Fig.~\ref{flapper}a and Fig.~\ref{flapper}b we have $\u ={\bf 0}$ on the boundary. For the symmetry plane $\theta=0$ in
Fig.~\ref{flapper}c we have the no-penetration conditions $u_\theta=0$, as well as  symmetry, $\p u_r/\p \theta =0$. 

The other boundary condition occuring on the flapper is the statement of its rotation, $\mathbf{u}(r,\theta=\alpha+\epsilon \Theta(t))=r\Omega(t)\e_\theta$ (with $\Omega(t) = \epsilon \dot{\Theta}$), which is obtained order by order by performing a  domain perturbation expansion. 
Consider a general vector field $\h(r,\theta)$. A Taylor expansion for $\h$ on the flapper position gives
\begin{equation}\label{TE}\h(r,\theta= \alpha + \epsilon \Theta)=\h(r,\alpha )+ \epsilon \Theta \frac{\p \h}{\p \theta}\bigg|_{(r,\alpha)} + ...,\end{equation}
which we apply  for $\h=\u - r\epsilon \dot{\Theta}\e_\theta={\bf 0}$, and therefore obtain, at order by order,
\begin{subeqnarray}\label{bc1}
\mathbf{u}_{1}&=&r \dot{\Theta} \langle \n \rangle, \\
\mathbf{u}_{2}&=&-\Theta\frac{\partial\mathbf{u_{1}}}{\partial\theta}-r \Theta\dot{\Theta}\langle \t \rangle,
\end{subeqnarray}
when evaluated  at $\theta = \alpha$.  In Fourier notations, we  use Eq.~\eqref{fourier} into Eq.~\eqref{bc1} and obtain the boundary conditions for the first two Fourier components as given by 
\begin{subeqnarray}
\tilde\u_1&=&i r\langle \n \rangle \slabel{bc_u1}, \\
\slabel{u20_bc}\tilde\u_2^{(0)}&=&-\frac{1}{2}\frac{\partial \tilde\u_1}{\partial\theta},
\slabel{u22_bc}
\end{subeqnarray}
evaluated at $\theta=\alpha$. As we see below, we do not need to solve for the Fourier component $\tilde\u_2^{(2)}$.

\subsection{Time-averaged stress on the flapper}
\label{averagestresses}

We are interested in calculating the time-averaged stress on the moving flapper. The unit normal into the fluid is  $ -\e_\theta$ and therefore  the instantaneous stress is given by 
\begin{equation}\label{stress}
\delta \f = -\sigma_{r\theta}\e_r - \sigma_{\theta\theta}\e_\theta.
\end{equation}
Using Taylor expansions for both the stress components and the direction vectors in Eq.~\eqref{stress} (i.e. applying Eq.~\eqref{TE} for $\h=-\bs\cdot\e_\theta$), we find  that there is no average stress at first order on the flapping surface, while 
the stress at order $O(\epsilon^2)$ 
along the averaged directions of the flapper, i.e. $(\langle {\bf t} \rangle,\langle {\bf n} \rangle )$, is given  by
\begin{subeqnarray}
\langle {\bf t} \rangle \cdot \delta\f_2 & = &- \sigma_{2,r\theta} - \Theta(t)\left(\frac{\p \sigma_{1,r\theta}}{\p \theta} -  \sigma_{1,\theta\theta}\right),\\
\langle {\bf n} \rangle \cdot \delta\f_2 & = &-\sigma_{2,\theta \theta} - \Theta(t) \left( \frac{\p \sigma_{1,\theta \theta}}{\p \theta}+ \sigma_{1,r\theta}\right), \quad
\end{subeqnarray}
evaluated at $\theta=\alpha$. In Fourier notations, and given that $\Theta(t)=\cos t$, they become
\begin{subeqnarray}\label{toget}
\langle {\bf t} \rangle \cdot \langle \delta \f_2 \rangle & = & -\Re\left\{\tilde \sigma_{2,r\theta} ^{(0)}
+ \frac{1}{2}\left(\frac{\p \tilde\sigma_{1,r\theta}}{\p \theta} -  \tilde\sigma_{1,\theta\theta}\right)\right\}, \quad \quad \\
\langle {\bf n} \rangle \cdot \langle  \delta\f_2  \rangle& = & -\Re\left\{\tilde \sigma_{2,\theta \theta} ^{(0)}
+ \frac{1}{2}  \left(\frac{\p \tilde\sigma_{1,\theta \theta}}{\p \theta}+ \tilde\sigma_{1,r\theta}\right)\right\},
\end{subeqnarray}
evaluated at $\theta=\alpha$. Thus, in order to compute the average stress at this order, we only need to know the Fourier components of the flow $\tilde\u_1$ and $\tilde\u_2^{(0)}$, and do not need to solve for the time-varying contribution given by $\tilde\u_2^{(2)}$.

Finally, on the no slip walls, there is no moving boundary. The average stress is therefore simply given there by the  Fourier component, $\tilde\bs_2^{(0)}$, evaluated on the wall.


\section{Single flapper}
\label{sec_single}

We now  solve for the case where there is only one flapper. 
As stated above, we solve the flow in the domain $0\leq \theta \leq \alpha$, and deduce the total stress by analogy.

\subsection{Solution at $O(\epsilon)$}

\subsubsection{Velocity field}

Taking the Fourier transform of Eq.~\eqref{order1}, we get $\nabla^4 \tilde\Psi_1 = 0 $. With the boundary conditions given by Eq.~\eqref{bc_u1},  we obtain the solution at first order
\begin{eqnarray}
\tilde\Psi_1&=&\frac{i r^{2}(-\sin2\theta+\tan\alpha\cos2\theta+2\theta-\tan\alpha)}{4(\tan\alpha-\alpha)}, \quad \quad
\end{eqnarray}
which is  the same as for the Stokes problem \cite{moffatt64}.

\subsubsection{Pressure}

The deviatotic stress is given by Eq.~\eqref{tau1_Fourier}, so we have
\begin{equation}\label{tau1}
\tilde\btau_{1}=\left(\frac{1+i\De_2}{1+i\De_1} \right)
\frac{i}{(\tan\alpha-\alpha)}{\bf T},
\end{equation}
with
\begin{subeqnarray}
 T(1,1)&=& 1-\cos2\theta-\tan\alpha\sin2\theta, \\ 
 T(1,2)&=&\sin2\theta-\tan\alpha\cos2\theta, \\
 T(2,1)&=&T(1,2),\\
 T(2,2)&=& -T(1,1).
\end{subeqnarray}
The pressure is found by integrating Cauchy's equation, Eq.~\eqref{cauchy_2}. We calculate the divergence of the deviatoric stress tensor, Eq.~\eqref{tau1}, and integrate it to find
\begin{equation}\label{p1}
\tilde p_1 = \left(\frac{1+i\De_2}{1+i\De_1} \right) \frac{2i}{\tan\alpha-\alpha} \ln \left(\frac{r}{r_0}\right),
\end{equation}
which has an integrable singularity at the origin ($r_0$ is a dimensionless microscopic cutoff length) \cite{moffatt64}.

\subsubsection{Contribution to second-order stress}
With the stress calculations above, we get at the average flapper position
\begin{subeqnarray}
\Re\left\{\frac{\p \tilde\sigma_{1,r\theta}}{\p \theta} -  \tilde\sigma_{1,\theta\theta}\right\} & = & 
\frac{\De_1-\De_2}{1+\De_1^2}  \left(\frac{2}{\tan\alpha-\alpha} \right) \quad \quad  \\
&& \times \left[ 1+\ln \left(\frac{r}{r_0}\right) \right],\nonumber \\
\Re\left\{ \frac{\p \tilde\sigma_{1,\theta \theta}}{\p \theta}+ \tilde\sigma_{1,r\theta}\right\} & = & 
\frac{\De_1-\De_2}{1+\De_1^2}  \left(\frac{\tan\alpha}{\alpha-\tan\alpha} \right),
\end{subeqnarray}
which, as expected, are equal to zero in the Newtonian limit,  $\De_1=\De_2$.

\subsection{Solution at $O(\epsilon^2)$}
\subsubsection{Velocity field}

\paragraph{Boundary conditions.}
The boundary conditions for the average flow at the location of the flapper, Eq.~\eqref{u20_bc}, become 
\begin{equation}\label{bc_mean2}
\tilde\u_2^{(0)}
=
\di \frac{-ir\alpha}{2(\tan\alpha-\alpha)}\langle {\bf t} \rangle\cdot 
\end{equation}

\paragraph{Average flow.}
Following Eq.~\eqref{order2_tocalculate}, we calculate
\begin{equation}
\tilde\u_1\cdot\nabla\tilde\gamd_1-\left(^{t}\nabla\tilde\u_1\cdot\tilde\gamd_1+\tilde\gamd_1\cdot\nabla\tilde\u_1\right)=\frac{\A}{\left(\tan\alpha-\alpha\right)^{2}}
\end{equation}
with 
\begin{subeqnarray}
 A(1,1)&=& 2+\tan^{2}\alpha-2\cos2\theta-2\tan\alpha\sin2\theta \quad \quad \\ 
 &&-\left(\tan\alpha-2\theta\right)\left(\sin2\theta-\tan\alpha\cos2\theta\right)\nonumber,\\
 A(1,2)&=&-\sin2\theta\left(1+\tan^{2}\alpha\right)\\ 
 && +2\theta(\cos2\theta+\tan\alpha\sin2\theta), \nonumber\\
 A(2,1)&=&A(1,2),\\
A(2,2)&=& 2+\tan^{2}\alpha-2\cos2\theta-2\tan\alpha\sin2\theta\\ 
&&+\left(\tan\alpha-2\theta\right)\left(\sin2\theta-\tan\alpha\cos2\theta\right)\nonumber.
\end{subeqnarray}
Since $\tilde \u_1$ is purely imaginary, the equation for the zeroth harmonics  becomes then
\begin{equation}
  \tilde\btau_2^{(0)}-\tilde\gamd_2^{(0)}
  =-\frac{1}{2\left(\tan\alpha-\alpha\right)^{2}}\frac{\text{De}_{2}-\text{De}_{1}}{1+i\text{De}_{1}}\A.
\end{equation}
In order to obtain the streamfunction at second order, we note that  $\nabla\cdot \A= {\bf 0}$ to get
\begin{equation}
\nabla\cdot \tilde\btau_2^{(0)}-\nabla\cdot\tilde\gamd_2^{(0)} = {\bf 0}.
\end{equation}
Combining with Cauchy's equation, Eq.~\eqref{cauchy_2} and taking the curl, we get 
\begin{equation}
\nabla^{4}\tilde\Psi_2^{(0)}=0.
\end{equation}
Given that the boundary condition, Eq.~\eqref{bc_mean2}, is purely imaginary, we obtain therefore the following result: there is not average flow at order $\epsilon^2$ ($\Psi_2^{(0)}=0$).

\subsubsection{Pressure and stress field}
We now have
\begin{equation}
\nabla\cdot\tilde\btau_2^{(0)}={\bf 0},
\end{equation}
and therefore the pressure at order $\epsilon^2$ is uniform (zero).

\subsubsection{Contribution to stress}

On the flapper ($\theta=\alpha$), we have
\begin{subeqnarray}
\tilde \sigma_{2,r\theta} ^{(0)} & = & \left(\frac{\De_1-\De_2}{1+i\De_1}\right)\frac{1}{\alpha-\tan \alpha},\\
\tilde \sigma_{2,\theta\theta} ^{(0)} & = &  \left(\frac{\De_1-\De_2}{1+i\De_1}\right)\frac{\tan \alpha}{\tan \alpha-\alpha},
\end{subeqnarray}
whereas on the  bottom wall ($\theta=0$) we obtain
\begin{subeqnarray}
\tilde \sigma_{2,r\theta} ^{(0)} (\theta=0) & = & 0,\\
\tilde \sigma_{2,\theta\theta} ^{(0)} (\theta=0)& = &  0.
\end{subeqnarray}

\subsection{Total time-averaged stress}
The calculations above allow us to evaluate Eq.~\eqref{toget} and obtain the time-averaged stress on the flapper at second order \begin{subeqnarray}\label{single}
\langle {\bf t} \rangle \cdot \langle \delta \f_2 \rangle & = & 
\frac{\De_2-\De_1}{1+\De_1^2}\left(\frac{\ln \left({r}/{r_0}\right)}{\tan \alpha-\alpha}\right),
\\
\langle {\bf n} \rangle \cdot \langle  \delta\f_2  \rangle& = & 
\frac{\De_2-\De_1}{2(1+\De_1^2)}\left(\frac{\tan \alpha}{\tan \alpha-\alpha}\right)\cdot
\end{subeqnarray}

\subsection{Flapper near a wall}

We first apply our results to the case illustrated in Fig.~\ref{flapper}a, that of a flapper near an infinite no-slip surface. 
We can assume  $\alpha<\pi/2$ without loss of generality.
Combining Eq.~\eqref{single} for angles $\alpha$ and $\pi-\alpha$ leads to the total time-averaged (dimensionless) stress on the flapper at second order
\begin{equation}\label{nearwall}
\langle\delta \f_2 \rangle = 
\frac{\pi\De(\eta_s/\eta-1)}{2(1+\De^2)}
\frac{\left[
\tan \alpha \langle \n\rangle + 2 \ln \left(\frac{r}{r_0}\right)
\langle \t\rangle 
\right]
}{(\tan\alpha-\alpha)(\pi + \tan \alpha - \alpha)}\cdot
\end{equation}
Note that in Eq.~\eqref{nearwall}, we have replaced the two Deborah numbers by using the definition $\De=\De_1$, and $\eta_s$ is the solvent viscosity, $\eta_s= \eta\lambda_2 /\lambda_1$ ($\eta_s\leq \eta$). Note also that  the stress  has an integrable singularity at the origin \cite{moffatt64}.

The main conclusion from Eq.~\eqref{nearwall} is that the scallop theorem breaks down for a viscoelastic fluid: Although the flapper is acting on the fluid with a reciprocal (time-periodic) motion, it is able to generate net  forces.  
Since $\eta_s \leq \eta$, we see that $\langle\delta \f_2 \rangle $ is directed along the $-\n$ and $-\t$ directions.
In the Newtonian limit, we have $\De =0$, and recover therefore the scallop theorem, $\langle\delta \f_2 \rangle ={\bf 0}$.
Physically, the quadratic forces in Eq.~\eqref{nearwall} stem from normal-stress differences which arise in polymeric fluids due to flow-induced stretching of polymer molecules \cite{birdvol1,birdvol2,doi88,larson99}.

\subsection{Flapper in a corner}
The formula above for a single flapper can be generalized when the flapper is located in a wedge-like geometry of size 
 $2\beta$ (Eq.~\ref{nearwall} is the particular case $\beta=\pi/2$). We can assume $\alpha \leq \beta$ without loss of generality. In that case the total  time-averaged stress along the flapper is given by
\begin{subeqnarray}\label{corner}
\langle \n\rangle \cdot \langle\delta \f_2 \rangle  & = &  \frac{\De(\eta_s/\eta-1)}{2(1+\De^2)}\times \\
&& \left[
\frac{\alpha\tan(2\beta-\alpha)+(\alpha-2\beta)\tan\alpha}{(\tan\alpha-\alpha)(\tan(2\beta-\alpha)-2\beta+\alpha)}
\right]\nonumber,
\\
\langle \t\rangle \cdot \langle\delta \f_2 \rangle   & = &  \frac{\De(\eta_s/\eta-1)}{1+\De^2}\ln \left(\frac{r}{r_0}\right)\times \\
&& \left[
\frac{\tan\alpha + \tan (2\beta-\alpha) - 2\beta }{(\tan\alpha-\alpha)(\tan(2\beta-\alpha)-2\beta+\alpha)}
\right] \nonumber\cdot
\end{subeqnarray}
The sign of the force in Eq.~\eqref{corner} is easily obtained numerically. In particular, by tuning the values of $\beta$ and $\alpha$,  it is straightforward to show that all four sign combinations for the components $\langle \n\rangle \cdot \langle\delta \f_2 \rangle  $ and $\langle \t\rangle \cdot \langle\delta \f_2 \rangle $ are possible.


\section{A scallop-like flapper}
\label{sec_scallop}

We examine here the case illustrated in Fig.~\ref{flapper}c  for a tethered scallop-like two dimensional flapper. The difference with the calculations above lies in the boundary conditions, as explained in \S\ref{bc}. We perform the calculations below for the domain  $0\leq \theta \leq \alpha$, and proceed to obtain the total time-averaged stress by analogy.

\subsection{Solution at $O(\epsilon)$}
\subsubsection{Velocity field}
With the appropriate boundary conditions at $\theta=0$, the solution at first order is given by
\begin{eqnarray}
\tilde\Psi_1&=&\frac{ir^{2}(\sin2\theta-2\theta\cos2\alpha)}{2(2\alpha\cos2\alpha-\sin2\alpha)}\cdot
\end{eqnarray}

\subsubsection{Pressure and stress field}
The first-order deviatoric stress tensor is given by
\begin{equation}
\tilde\btau_{1}=
\frac{1+i\De_2}{1+i\De_1}
\frac{2i}{(2\alpha\cos2\alpha-\sin2\alpha)}
{\bf T},
\end{equation}
with
\begin{subeqnarray}
T(1,1)&=& \cos2\theta-\cos2\alpha, \\ 
T(1,2)&=& - \sin2\theta,\\
T(2,1)&=& T(1,2),\\
T(2,2)&=& -T(1,1).
\end{subeqnarray}
The  divergence of $\tilde\btau_{1}$ can be integrated to lead to the pressure
\begin{equation}
\tilde p_1 = \frac{1+i\De_2}{1+i\De_1} \left(\frac{4i}{\tan 2\alpha - 2\alpha}\right)\ln\left(\frac{r}{r_0}\right).
\end{equation}

\subsubsection{Contribution to second-order stress}
As part of Eq.~\eqref{toget}, we have the following
\begin{subeqnarray}\label{part1}
\Re\left\{\frac{\p \tilde\sigma_{1,r\theta}}{\p \theta} -  \tilde\sigma_{1,\theta\theta}\right\} & = & 
\frac{(\De_1-\De_2)}{1+\De_1^2} \frac{4\left[1+\ln\left({r}/{r_0}\right)\right] }{\tan 2\alpha-2\alpha}, \quad\quad  \\
\Re\left\{ \frac{\p \tilde\sigma_{1,\theta \theta}}{\p \theta}+ \tilde\sigma_{1,r\theta}\right\} & = & 
\frac{(\De_1-\De_2)}{1+\De_1^2}\frac{2\tan 2\alpha}{2\alpha-\tan 2\alpha},
\end{subeqnarray}

\subsection{Solution at $O(\epsilon^2)$}
\subsubsection{Velocity field}
\paragraph{Boundary conditions.}
Here, the  boundary conditions from Eq.~\eqref{u20_bc} become 
\begin{equation}
\tilde\u_2^{(0)}
=\frac{ir}{2} \frac{2\alpha + \tan2\alpha}{2\alpha -  \tan2\alpha}\langle {\bf t}\rangle\cdot
\end{equation}
\paragraph{Average flow.}
After some algebra we get
\begin{eqnarray}
&& \tilde\u_1\cdot\nabla\tilde\gamd_1-\left(^{t}\nabla\tilde\u_1\cdot\tilde\gamd_1+\tilde\gamd_1\cdot\nabla\tilde\u_1\right)\\
&&\quad \quad =\frac{4\B}{\left(2\alpha\cos2\alpha-\sin2\alpha\right)^{2}},
\end{eqnarray}
with
\begin{subeqnarray}
B(1,1)&=& 1+\cos2\alpha (2\theta\sin2\theta + \cos2\alpha - 2 \cos 2\theta) \\ 
B(1,2)&=& \cos 2\alpha (2\theta\cos2\theta-\sin2\theta),\\ 
B(2,1)&=& B(1,2),\\
B(2,2)&=& 1+\cos2\alpha ( \cos2\alpha - 2 \cos 2\theta-2\theta\sin2\theta). \quad \quad
\end{subeqnarray}
As in the case of a single flapper, we observe that $\nabla\cdot \B = {\bf 0}$ to obtain that there is no average flow at second order, $\Psi_2^{(0)}=0$.

\subsubsection{Pressure and stress field}
We have
\begin{equation}
\tilde \btau_2^{(0)}=\frac{2(\De_1-\De_2)}{1+i\De_1} 
\frac{\B}{\left(2\alpha\cos2\alpha-\sin2\alpha\right)^{2}},
\end{equation}
as well as zero pressure ($p_2 =0$), so that the average stress tensor on the flapper is given by
\begin{subeqnarray}\label{part2}
\tilde \sigma_{2,r\theta} ^{(0)} & = & \frac{2(\De_1-\De_2)}{1+i\De_1} \frac{1}{2\alpha-\tan2\alpha},\\
\tilde \sigma_{2,\theta\theta} ^{(0)} & = &  
\frac{2(\De_1-\De_2)}{1+i\De_1}  
\frac{\tan 2\alpha}{\tan 2\alpha-2\alpha }\cdot
\end{subeqnarray}

\subsection{Total time-averaged stress}
With the results of Eqs.~\eqref{part1} and \eqref{part2}, we obtain the time-averaged stress on the flapper as given by
\begin{subeqnarray}
\langle {\bf t} \rangle \cdot \langle \delta \f_2 \rangle & = & \frac{2(\De_2-\De_1)}{1+\De_1^2}\frac{\ln\left({r}/{r_0}\right)}{\tan2\alpha-2\alpha},\\
\langle {\bf n} \rangle \cdot \langle  \delta\f_2  \rangle& = & 
\frac{\De_2-\De_1}{1+\De_1^2}\frac{\tan2\alpha}{\tan2\alpha-2\alpha}\cdot
\end{subeqnarray}

\subsection{Time-averaged stress on scallop}
Once we have the calculation for flapper in the domain $0\leq \theta \leq \alpha$, we can use two reflection symmetries to obtain the time-averaged  stress on the entire scallop. We get
\begin{eqnarray}\label{scallopforce}
\langle\delta \f_2 \rangle  &=&  \frac{(\eta_s/\eta-1)\De}{1+\De^2}
 \times \\
&& \frac{ 4\pi\cos\alpha\left[2\ln\left({r}/{r_0}\right) - \tan\alpha \tan 2\alpha
\right]}
{(\tan 2\alpha - 2\alpha)(2\pi+\tan 2\alpha - 2\alpha)}
\e_x \nonumber\cdot
\end{eqnarray}
Interestingly, in this simplified two-dimensional setup, the sign of the  net stress can only be established when $\pi/4\leq \alpha \leq \pi/2$. Indeed, when $0\leq \alpha \leq \pi/4$, the numerator of Eq.~\eqref{scallopforce} has an undetermined sign and depends on the size of the scallop. If we denote by $\alpha_c$ the solution of $\tan 2\alpha_c = 2(\alpha_c-\pi)$ ($\alpha_c\approx 51.27^\circ$), the  force is in the $-x$ direction for  $\alpha_c \leq \alpha \leq \pi/2$ and in the $+x$ direction for 
$\pi/4\leq \alpha \leq \alpha_c$.

\section{Discussion}
\label{discussion}

\subsection{Viscoelastic propulsion.} 

In this paper, we have used asymptotic calculations to derive the time-averaged forces acting on a tethered flapper in a polymeric (Oldroyd-B) fluid.  The main results we obtained are Eqs.~\eqref{nearwall}, \eqref{corner} and \eqref{scallopforce}, which give the dimensionless stress acting on a flapper near  a wall, in a wedge, and on a two-dimensional scallop respectively. 

The calculations presented above were made under a number of simplifying assumptions. First and foremost, we have assumed the flapper to be tethered, and therefore do not solve for the swimming motion that would be caused  by the net forces it is generating. The second simplifying assumption is that of a two-dimensional geometry. In that regard our work is the extension of classical work on similarity solutions for viscous corner flows \cite{taylor_scraper, batchelor_book,moffatt00}. The advantage of considering such an idealized geometrical setting is to allow us to solve rigorously the equations of motion for the viscoelastic flow, and obtain in a closed form the resulting time-averaged forces.

Within the framework delimited by these assumptions, our results show rigorously that, using a purely harmonic motion, it is possible to escape the constraints of the scallop theorem in a viscoelastic fluid, even in the absence of inertia. 
Since the flapper is  moving with a pure sinusoidal motion, the forces arising on the flappers do not originate from any  rate-dependent mechanisms  (which could be the case, say, if it was opening fast, and closing slowly). Instead, the net force arises from a rectification of  the time-periodic actuation by normal-stress differences ({\it i.e.} flow-induced tension along streamlines due to the stretching of polymeric molecules). We note that since the flow is not viscometric, the explicit relationship between classically measured rheological parameters (such as normal stress coefficients) and the magnitude and direction of the time-averaged forces is not evident. 
A similar situation - but simpler to analyze -  arises in a cone-and-plate rheometer  where the flow is viscometric. Under the oscillatory motion of the rheometer  quadratic  forces proportional to the first normal stress coefficient of the fluid act to separate the cone from the plate \cite{birdvol1}. These are the same forces which are responsible for the famous non-Newtonian rod-climbing effect  \cite{birdvol1}.

\subsection{Order of magnitude and biological relevance.}

In our work, the hinge point of the  flapper is fixed in place. The force arising from the fluid correspond therefore to tethered swimming, and are directly relevant to the motion of cilia-like biological appendages. In particular, our results show that, as a difference with a Newtonian fluid,  the component in the motion of a cilium which is reciprocal (its periodic  back and forth motion) does contribute to force generation in polymeric liquids (e.g. airway mucus.)

In previous work, it was shown that  viscoelastic stress systematically lead to a decrease of the swimming performance for swimmers who deform their flagella in  a wave-like fashion \cite{lauga07,Fu08,Fu_swimming}. The calculations above display explicitly a class of body deformation where stresses arising from complex fluids act the other way, {\it i.e.} actually increase the forces generated. In the case of a general  body deformation, both are in general possible and the beneficial vs. detrimental nature of viscoelastic stresses cannot be established a priori. It is also worth pointing out that if the flapper was flexible, it would deform as a result of fluid forces, and the resulting non-reciprocal shapes would lead to force generation even in the Newtonian limit  \cite{machin58,WigginsGoldstein}.

Let us now estimate the order of magnitude of the viscoelastic forces reported here, and consider for simplicity the case of a single flapper near a wall (Fig.~\ref{flapper}a). In that case, the result of Eqs.~\eqref{nearwall} leads to the order of magnitude estimate for the time-averaged force
\begin{equation}\label{scaling}
|\langle \delta {\bf f}_2\rangle |\sim\epsilon^2  \frac{\lambda \omega}{1+(\lambda \omega)^2}\left(1-\frac{\eta_s}{\eta}\right) \eta \omega .
\end{equation}
As an example of  biological polymeric fluid, we consider cervical mucus \cite{wolf77_1,wolf77_2,wolf77_3,wolf78,wolf80,tam80}. Since the viscosity of cervical mucus is at least two orders of magnitude that of water, we see that a large-amplitude flapping motion ($\epsilon\sim 1$) at frequency $\omega \sim 1/\lambda$ leads to $|\langle \delta {\bf f}_2\rangle |\sim\eta \omega$. Remarkably, stresses on the order of those that would be generated in a purely viscous fluid with viscosity $\eta$ and sheared at a rate $\omega$ can be generated {\it on average} inside the polymeric fluid. For cervical mucus, we have $\eta\sim 10^{-1}-10$ Pas. Since $\lambda\sim 1-100$ we would get $\omega\sim 10^{-2}-1$ s$^{-1}$, leading to stresses on the order $|\langle \delta {\bf f}_2\rangle |\sim 10^{-3}-10$ Pa, similar to the range of stresses applied for example by a ciliary array on an overlying water-like fluid.  The  effect reported in our paper could be therefore  significant in a biological setting.

A second important feature is to estimate how easy it would be to measure these viscoelastic forces in an experiment. Indeed, the forces on the flapper also include an unsteady component, which is of order $\epsilon$, and (although it averages to zero) might dominate the time-averaged force during experimental measurements. The amplitude of this unsteady  term, in dimensional form, can be  inferred from Eqs.~\eqref{tau1}-\eqref{p1}, and we see that
\begin{equation}
|\delta {\bf f}_1 |\sim \epsilon \left(\frac{1+(\lambda \omega)^2({\eta_s}/{\eta})^2}{1+(\lambda \omega)^2}\right)^{1/2}\eta \omega.
\end{equation}
When $\lambda \omega \sim 1$, and $\epsilon\sim 1$ we obtain the estimate  $|\delta {\bf f}_1 |\sim \eta \omega$. Therefore for large-amplitude motion and a flapping rate comparable to the inverse relaxation time of the fluid, the unsteady forces on the flapper are expected to be  of the same order as their time- average $|\delta {\bf f}_1 |\sim|\langle \delta {\bf f}_2\rangle |\sim\eta\omega$. In that case, the breakdown of the scallop theorem would be significant, and  could realistically be measured in an experiment.

Finally, let us address the implication of our results to the locomotion of microorganisms. Our calculations assume the swimmer to be tethered, and therefore the analysis would have to be carried out separately in the case of free-swimming. This would also require to consider a finite-size body. Qualitatively, the time-averaged forces obtained above would lead to propulsion. The organism would swim at a speed found by balancing the propulsive forces, Eq.~\eqref{scaling}, with a viscoelastic drag. In theory, given that swimmers with arbitrarily complex geometries could be devised, locomotion could occur with arbitrarily complex trajectories. A systematic study of viscoelastic swimming for finite-size bodies will be reported in future work.

\subsection{Generalization to other fluids.}

Our calculations were performed for an Oldroyd-B fluid, which, in steady-shear, has a constant viscosity and first normal stress coefficient, and no second-normal stress differences. In previous work on the swimming of a sheet in a viscoelastic fluid  \cite{lauga07}, the Oldroyd-B results were extended  to a more general class of Oldroyd-like models (Johnson-Segalman-Oldroyd), to the (nonlinear) Giesekus model, both of which display  shear-thinning behavior for the  viscosity and both normal stress coefficients in steady shear. A more realistic polymeric model was also considered, FENE-P, which remains physically accurate for large values of the Deborah numbers. For the FENE-P model, the viscosity and first normal stress coefficient are shear-thinning, and the second normal stress coefficient is zero.

Following the calculations in the appendices of Ref.~\cite{lauga07}, it is straightforward to extend the  flapping calculations above  to these models. Our results, Eqs.~\eqref{nearwall}, \eqref{corner} and \eqref{scallopforce}, turn out to be  unchanged (not reproduced here), and therefore appear to be quite general.

\subsection{Force vs. flow generation.}

An interesting feature of our asymptotic results is that the net  forces occur at order $O(\epsilon^2)$. However, there is no time-averaged flow of the polymeric fluid at this order. The same result is obtained for all the constitutive models considered. A detailed (but lengthy) analysis reveals that a time-averaged flow actually occurs at order $O(\epsilon^4)$, and will be reported in future work. How can we reconcile the fact that there could be a net force without a flow? In fact,  a similar situation arises in the example of the  cone-and-plate rheometer proposed above. Under the oscillatory motion of the rheometer, not net flow is generated, but time-averaged forces are arising.

\subsection{Asymptotic details.} 
As in the classic work by  HK Moffatt \cite{moffatt64}, the calculations here should be thought of as similarity solution, valid  away from the microscopic cuttoff length, $r_0$ at which the pressure divergence is regularized, but close enough to the hinge point that inertial effects can be neglected, {\it i.e.}  $\epsilon \omega r^2/\nu\ll 1$ or $r\ll\sqrt{\nu/\epsilon \omega}$, as well as edge effects for finite-size flappers.

\section*{Acknowledgments}
We thank Henry Fu and Tom Powers for useful discussions. This work was supported in part by the National Science Foundation (Grants No. CTS-0624830 and CBET-0746285).

\bibliographystyle{unsrt}
\bibliography{flapper_citations}

\end{document}